\documentclass[aps,prl,twocolumn,letterpaper,superscriptaddress,showpacs,nobibnotes]{revtex4-2}
\usepackage{keyval}%
\usepackage{graphicx}
\usepackage{dcolumn}
\usepackage{bm}
\usepackage{color}
\usepackage{amsmath,amssymb}

\begin{document}

\title{Disorder-Mediated Magnetism in Transition-Metal Dichalcogenide Monolayers: XS$_2$ and XSe$_2$ (X = Ti, V, Co, Mo, and Ru)}
\author{Chi-Cheng Lee}
%\email{Email: cclee.physics@gmail.com}
\affiliation{Department of Physics, Tamkang University, Tamsui, New Taipei 251301, Taiwan}%
\author{Yukiko Yamada-Takamura}
\affiliation{School of Materials Science, Japan Advanced Institute of Science and Technology (JAIST), 1-1 Asahidai, Nomi, Ishikawa 923-1292, Japan}%
\author{Taisuke Ozaki}
\affiliation{Institute for Solid State Physics, The University of Tokyo, 5-1-5 Kashiwanoha, Kashiwa, Chiba 277-8581, Japan}%
\date{\today}

\begin{abstract}
Magnetism in transition-metal dichalcogenide monolayers has attracted considerable attention, yet its nature remains elusive, particularly in monolayer 1T-VSe$_2$. Our first-principles calculations show that Stoner instability can drive ferromagnetism in XS$_2$ and XSe$_2$ monolayers (X = Ti, V, Co, Mo, and Ru), associated with a reduced density of states at the Fermi level. We demonstrate that mixing nonmagnetic Ti, Co, Mo, and Ru disulfides or diselenides can induce magnetism. Conversely, mixing magnetic VS$_2$ or VSe$_2$ with other compounds can completely suppress magnetism. Both scenarios can lower the total energy. Moreover, substituting a small number of V atoms with Co defects can also disrupt ferromagnetism in 1T-V(S,Se)$_2$. Even in pristine 1T-VSe$_2$, we find ferrimagnetic configurations with reversed local V moments that are nearly degenerate with the ferromagnetic state. Their competition may contribute to the suppression of long-range ferromagnetic order. This reduced energy penalty for magnetic and structural disorder arises from a suppressed density of states at the Fermi level, resembling the mechanism underlying the Peierls transition. Electronically driven energy lowering near the Fermi level may therefore be widespread among transition-metal dichalcogenide monolayers in various forms.
\end{abstract}
  
\maketitle

The emergence of two-dimensional transition-metal dichalcogenides (TMDs) has provided versatile platforms for developing advanced electronic, spintronic, and optoelectronic technologies, as well as a broad range of other promising applications\cite{Manzeli}. Among these materials, MoS$_2$ is a prototypical compound that exhibits a thickness-dependent transition between direct and indirect band gaps\cite{Mak} and can host the quantum spin Hall effect\cite{Qian}. Owing to weak interlayer van der Waals interactions, monolayer phases of MoS$_2$, most notably the 2H and distorted 1T structures, have been experimentally realized\cite{Lin,Lina,Lei}. These structures, along with similar ones, are also found in other TMDs. Magnetism in layered TMDs has attracted considerable interest for its intriguing phenomena. For example, ferromagnetism can be induced in nonmagnetic monolayer MoS$_2$ by tuning the electron density\cite{Roch}. In bilayer VS$_2$, strong magnetoelectric coupling between ferroelectric polarization and antiferromagnetic order has been theoretically predicted\cite{Liu}, enabling electronic control of magnetism. Magnetic moments can also be introduced via substitutional impurities, such as Fe and V atoms occupying S sites on the MoS$_2$ surface\cite{Fuhr}. Topological magnetic objects, including skyrmions and antiferromagnetic Néel merons, have also been explored in CrTe$_2$-based heterobilayers\cite{Abuawwad}.

Among TMDs, the nature of magnetism in 1T-VSe$_2$ remains particularly elusive. Density functional theory (DFT) calculations within the generalized gradient approximation (GGA)\cite{GGA} predict a ferromagnetic ground state, though this prediction can vary with different approximations\cite{Ma,Li,Fuh,Esters,Yin}. Early experiments supported the GGA prediction, reporting ferromagnetic order even above room temperature\cite{Bonilla}, whereas subsequent studies found no evidence of long-range ferromagnetic order\cite{Wong, Feng,Fumega,Sumida}. Most experiments suggest that charge-density-wave formation competes with magnetism, potentially explaining the absence of ferromagnetism. Nevertheless, experiments show no ferromagnetism down to 2 K\cite{Wong}, 10 K\cite{Feng}, 5 K\cite{Fumega}, and 6 K\cite{Sumida}. Evidence of spin frustration and short-range magnetic interactions further suggests that ferromagnetism and antiferromagnetism may coexist\cite{Wong,Sumida}. Another issue is that DFT calculations constrained to nonmagnetic states with suppressed magnetism often reproduce the experimentally observed phases\cite{Feng,Fumega,Sumida}, despite the calculated energetic preference for the ferromagnetic state. These findings underscore the need to identify the physical ingredients missing from DFT-GGA for 1T-VSe$_2$.

In this study, we investigate magnetism in TMDs from first principles, with a focus on elucidating its nature in 1T-VSe$_2$. DFT-GGA calculations were performed using OpenMX\cite{Openmx,Ozaki}. Structures were visualized using VESTA\cite{Vesta}. All monolayer structures derived from the 2H and 1T phases, including distorted structures, are denoted as 1H and 1T, respectively. Phonon dispersions were computed from partitioned force constants obtained via supercell calculations\cite{Chi}. Further computational details and the phonon dispersions investigated are provided in the Supplemental Material\cite{Supplement}.

\begin{figure}[tbp]
\includegraphics[width=1.00\columnwidth,clip=true,angle=0]{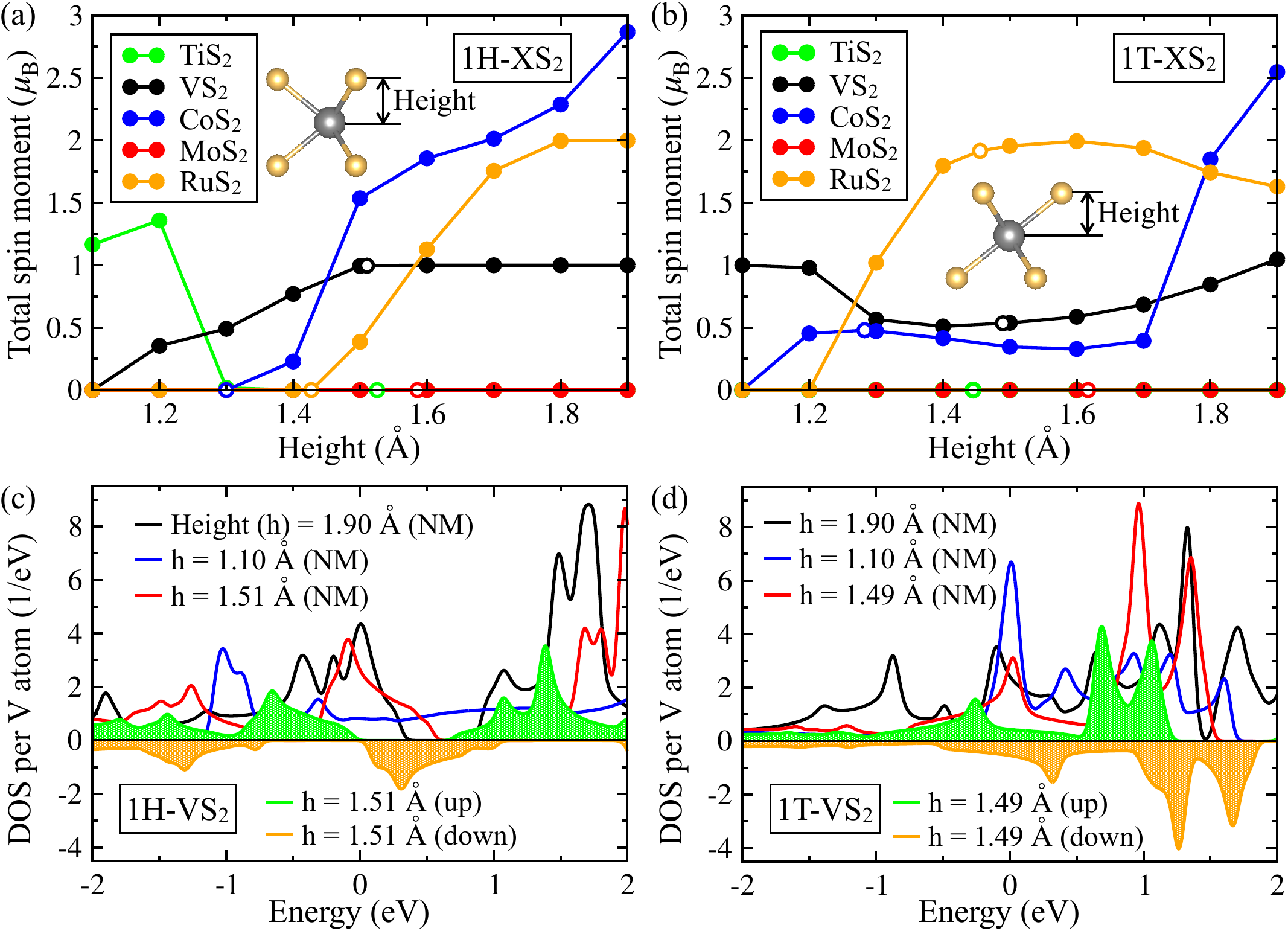}
\caption{Dependence of the total spin moment on S-atom height in (a) 1H- and (b) 1T-XS$_2$ for X = Ti, V, Co, Mo, and Ru, with lattice constants fixed at the values obtained from the fully relaxed structures. Open circles mark the results at the relaxed heights. The inset defines the S-atom height. Nonmagnetic (NM) density of states (DOS) of the V atom at selected S-atom heights in (c) 1H- and (d) 1T-VS$_2$, together with spin-polarized DOS at the relaxed heights, decomposed into up- (positive) and down-spin (negative) contributions.
}
\label{fig:fig1}
\end{figure}

Ferromagnetism can arise via the Stoner instability, in which exchange-induced splitting of spin-up and spin-down states overcomes the increase in band energy\cite{Stoner}. Consequently, energy competition near the Fermi level plays a key role, and the onset of ferromagnetic order generally requires a high density of states (DOS) at the Fermi level. In TMDs, the DOS is expected to be tunable by varying the chalcogen heights, which in turn affects ferromagnetism. First, we consider 1H-XS$_2$ compounds in their fully relaxed structures. 1H-TiS$_2$ and 1H-MoS$_2$ are nonmagnetic, whereas 1H-CoS$_2$ and 1H-RuS$_2$ lie near the boundary between nonmagnetic and ferromagnetic order. 1H-VS$_2$ is ferromagnetic. As shown in Fig.~\ref{fig:fig1}(a), the total spin moments of 1H-TiS$_2$, 1H-VS$_2$, 1H-CoS$_2$, and 1H-RuS$_2$ can be readily modulated by varying the S heights over a wide range, whereas 1H-MoS$_2$ is insensitive to this structural parameter. Among the 1T-XS$_2$ compounds, fully relaxed 1T-VS$_2$, 1T-CoS$_2$, and 1T-RuS$_2$ are ferromagnetic, with spin moments that also depend on the S heights, as shown in Fig.~\ref{fig:fig1}(b). By contrast, 1T-TiS$_2$ and 1T-MoS$_2$ remain nonmagnetic throughout the investigated S heights.

\begin{figure}[tbp]
\includegraphics[width=1.00\columnwidth,clip=true,angle=0]{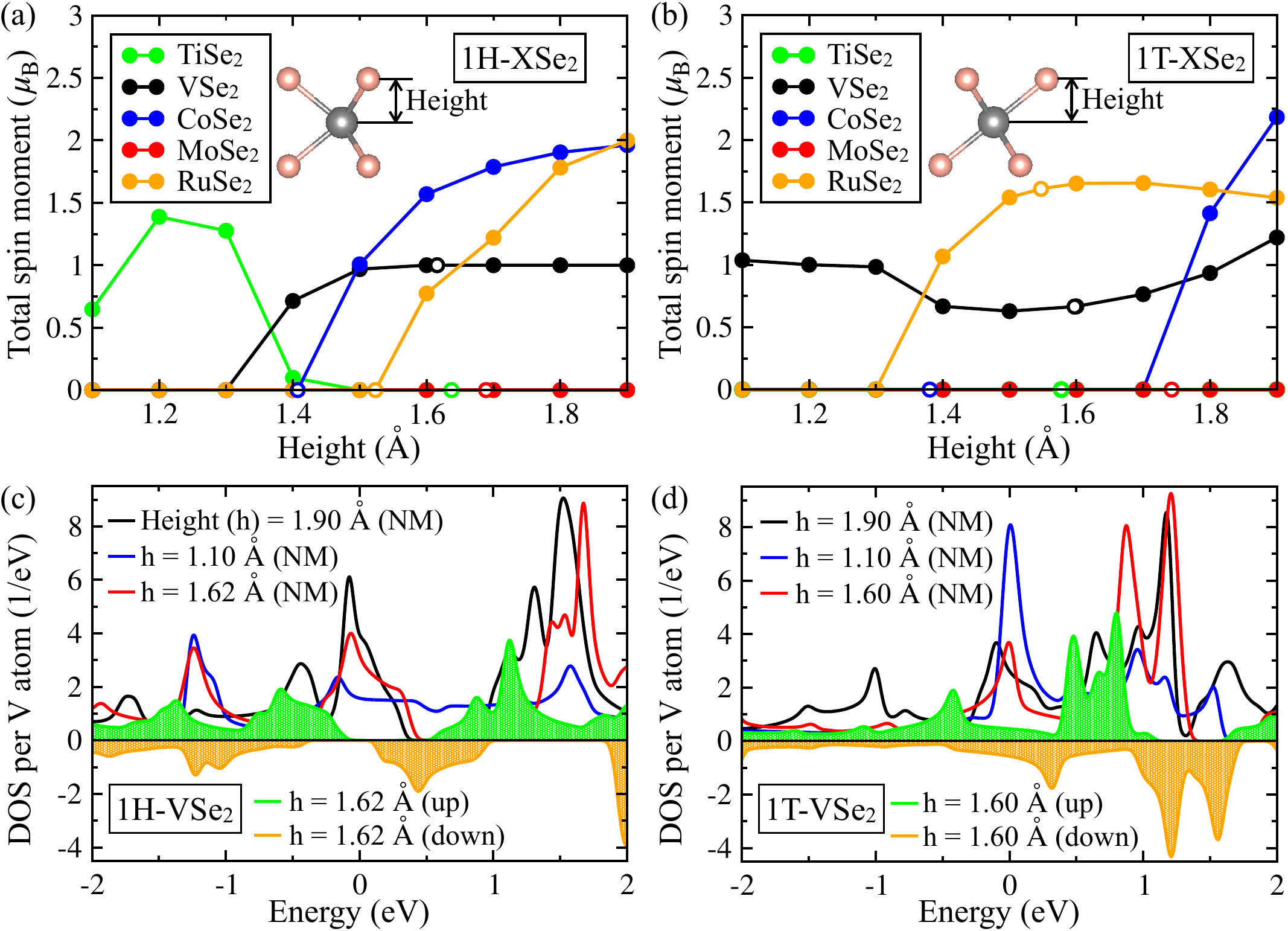}
\caption{Dependence of the total spin moment on Se-atom height in (a) 1H- and (b) 1T-XSe$_2$ for X = Ti, V, Co, Mo, and Ru, with lattice constants fixed at the values obtained from the fully relaxed structures. Open circles mark the total spin moments at the relaxed heights. The inset defines the Se-atom height. Nonmagnetic (NM) density of states (DOS) of the V atom at selected Se-atom heights in (c) 1H- and (d) 1T-VSe$_2$, together with spin-polarized DOS at the relaxed heights, decomposed into up- and down-spin contributions.
}
\label{fig:fig2}
\end{figure}

To clarify the relationship between the DOS at the Fermi level and the emergence of ferromagnetism in XS$_2$ monolayers, we present the spin-unpolarized DOS of the representative compounds 1H- and 1T-VS$_2$ at selected S heights in Figs.~\ref{fig:fig1}(c) and (d), respectively. A pronounced peak at the Fermi level (0 eV) appears at the S height where VS$_2$ exhibits ferromagnetism. By contrast, nonmagnetic 1H-VS$_2$ at an S height of 1.10 Å shows a relatively low DOS at the Fermi level. The spin-resolved DOS of the fully relaxed structures is also shown, revealing prominent splitting at the Fermi level between the spin-up and spin-down contributions. In 1H-VS$_2$, the spin-up and spin-down components are well separated, whereas in 1T-VS$_2$ the splitting is incomplete. These features are consistent with the Stoner instability scenario, and in VS$_2$ the energy lowering can be attributed to a reduction in the spin-polarized band energy due to splitting of the DOS at the Fermi level. A similar dependence of ferromagnetism on chalcogen height is observed in both 1H- and 1T-XSe$_2$, as shown in Figs.~\ref{fig:fig2}(a) and (b). The major differences from XS$_2$ are that ferromagnetism emerges at a modified chalcogen height, and fully relaxed 1T-CoSe$_2$ remains nonmagnetic. Similar DOS signatures are also observed in 1H- and 1T-XSe$_2$, as shown in Figs.~\ref{fig:fig2} (c) and (d), respectively. These results show how Stoner instability can drive ferromagnetism and lower the electronic energy in X(S,Se)$_2$ monolayers.

\begin{table}[tbp]
\caption{The relative total energies of fully relaxed monolayer 1H and 1T X(S,Se)$_2$ phases, calculated in (1$\times$1) and (6$\times$6) supercells, are reported in meV/atom. The mixing and relative energies of selected compounds are provided. E$_{\mathrm{mixing}}$ is defined as the energy of the selected phase relative to the total energy of the corresponding unmixed compounds in the same structural form, determined solely from the column labeled 1H-(6$\times$6) or 1T-(6$\times$6). E$_{\mathrm{relative}}$ is defined as the energy of the selected phase relative to the total energy of the corresponding unmixed compounds in their respective lower-energy structural forms, which may be either 1H or 1T.
}
\label{table:totalenergy}%
\begin{tabular}{ccccc}
\hline\hline
 X(S,Se)$_2$ & 1H-(1$\times$1) & 1H-(6$\times$6) & 1T-(1$\times$1) & 1T-(6$\times$6) \\ \hline
   CoS$_2$   &   165.6         &   110.9         &   36.5          &   0.0           \\
   MoS$_2$   &     0.0         &   0.0           &   280.6         &   194.2         \\
   RuS$_2$   &   249.0         &   164.9         &   227.1         &   0.0           \\
   TiS$_2$   &   137.2         &   133.4         &   2.9           &   0.0           \\
    VS$_2$   &   0.0           &   0.0           &   20.6          &   20.6          \\ \hline
  CoSe$_2$   &    142.0        &   112.8         &   23.2          &   0.0           \\
  MoSe$_2$   &    0.0          &   0.0           &   232.9         &   118.1         \\
  RuSe$_2$   &    231.1        &   132.8         &   222.6         &   0.0           \\
  TiSe$_2$   &    114.0        &   113.5         &   2.2           &   0.0           \\
   VSe$_2$   &    0.1          &   0.0           &   17.0          &   17.0          \\ \hline
             &                 &                                          &    E$_{\mathrm{mixing}}$    &  E$_{\mathrm{relative}}$    \\ \hline
\multicolumn{3}{c}{1H-Ti$_{0.25}$Co$_{0.25}$Mo$_{0.25}$Ru$_{0.25}$S$_2$}  &     -55.9          &    46.4            \\
\multicolumn{3}{c}{1H-Ti$_{0.25}$Co$_{0.25}$Mo$_{0.25}$Ru$_{0.25}$Se$_2$} &     -54.1          &    35.7            \\
\multicolumn{3}{c}{1T-Ti$_{0.25}$V$_{0.25}$Co$_{0.25}$Ru$_{0.25}$S$_2$}   &     -20.2          &    -15.1           \\
\multicolumn{3}{c}{1T-Ti$_{0.25}$V$_{0.25}$Co$_{0.25}$Ru$_{0.25}$Se$_2$}  &     -29.1          &    -24.9           \\
\multicolumn{3}{c}{1T-V$_{0.89}$Co$_{0.11}$S$_2$}                         &     -5.4           &    12.9            \\
\multicolumn{3}{c}{1T-V$_{0.89}$Co$_{0.11}$Se$_2$}                        &     -4.0           &    11.2            \\
\multicolumn{3}{c}{Ferrimagnetic (6$\times$6) 1T-VSe$_2$}                 &     -0.2           &    16.8            \\
\multicolumn{3}{c}{Ferrimagnetic ($3\sqrt{3}\times\sqrt{7}$) 1T-VSe$_2$}       &     -0.2           &    16.8            \\ \hline\hline
\end{tabular}%
\end{table}

Strong competition between magnetism and charge density waves (CDWs) is ubiquitous in TMDs. This competition can be understood within the Peierls picture, in which CDW formation opens a gap at the Fermi level by coupling electronic states connected by the relevant wave vectors. The resulting suppression of the DOS at the Fermi level prevents the Stoner criterion for magnetism. Table~\ref{table:totalenergy} reports the total energies of structures obtained by introducing small structural perturbations into the (6$\times$6) supercells of the 1H and 1T phases of X(S,Se)$_2$ and subsequently performing full structural relaxation. Within each row, energies are reported relative to the lowest-energy configuration. Except for 1H-Mo(S,Se)$_2$ and 1H- and 1T-V(S,Se)$_2$, all investigated X(S,Se)$_2$ systems lower their total energies by forming CDWs in the (6$\times$6) supercells. We have also examined phonon instabilities and found that (1$\times$1) 1H-Mo(S,Se)$_2$, (1$\times$1) 1H- and 1T-V(S,Se)$_2$, and all (6$\times$6) X(S,Se)$_2$ are dynamically stable, with no imaginary frequencies\cite{Supplement}. For 1H-VSe$_2$, the energy reduction in the (6$\times$6) supercell is only 0.1 meV/atom, which is within our computational uncertainty; the relaxed structure remains essentially identical to that of (1$\times$1) 1H-VSe$_2$. After examining magnetism, we find that all explored (6$\times$6) X(S,Se)$_2$ are non-magnetic except 1H- and 1T-V(S,Se)$_2$.

\begin{figure}[tbp]
\includegraphics[width=1.00\columnwidth,clip=true,angle=0]{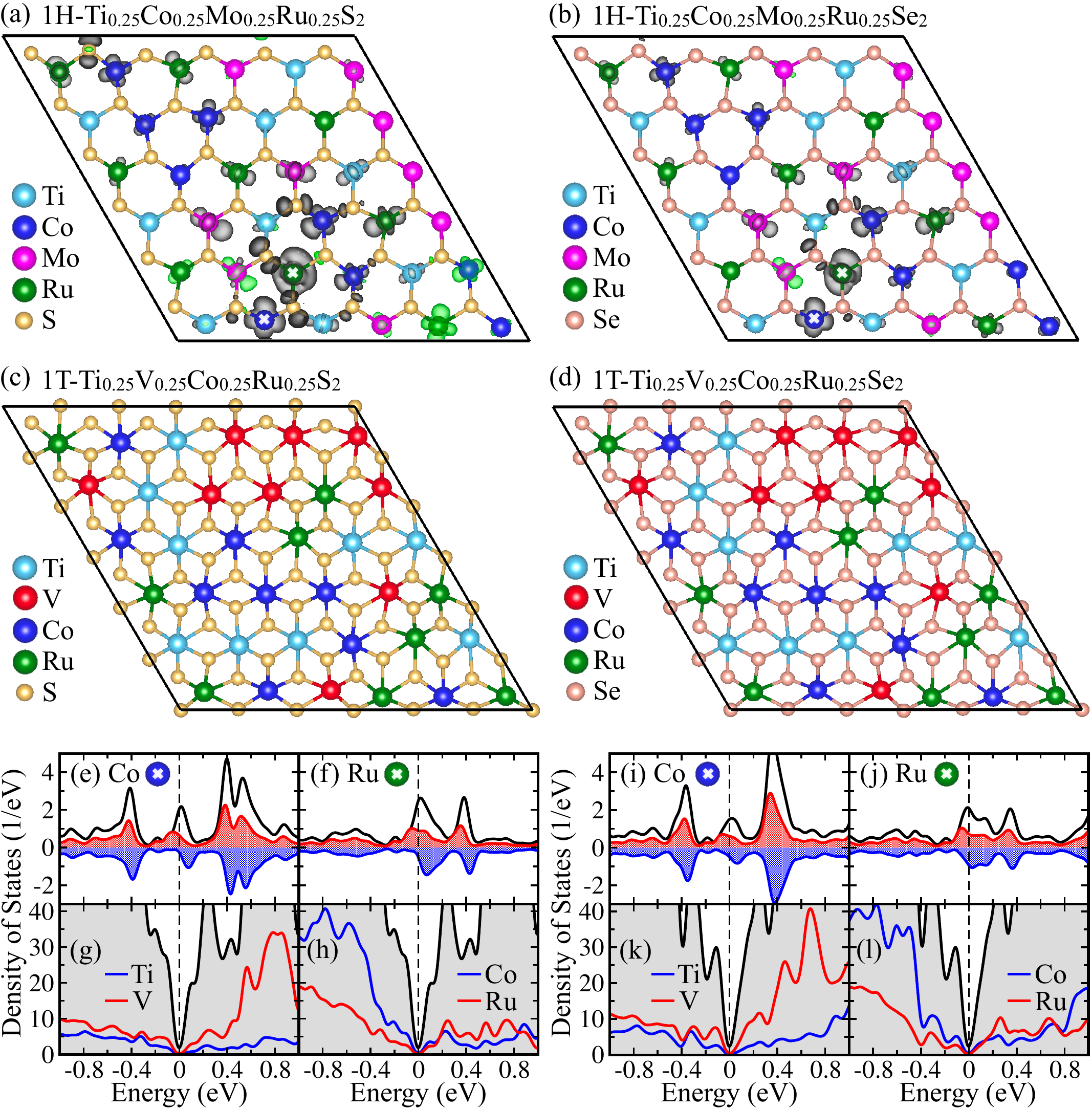}
\caption{Spin-density isosurfaces at 0.0005 electron/Bohr$^3$ for 1H phases mixing Ti, Co, Mo, and Ru (a) disulfides and (b) diselenides, modeled with a $6\times6$ supercell. Positive and negative spin densities are shown in black and green, respectively. Structures of (c) 1T-Ti$_{0.25}$V$_{0.25}$Co$_{0.25}$Ru$_{0.25}$S$_{2}$ and (d) 1T-Ti$_{0.25}$V$_{0.25}$Co$_{0.25}$Ru$_{0.25}$Se$_{2}$, neither of which exhibits spin-polarized density. Spin-up (red) and spin-down (blue) contributions in the density of states (DOS) for the selected (e) Co and (f) Ru atoms indicated in (a), together with the nonmagnetic DOS (black curves) calculated using the same structure. DOS contributions from (g) Ti and V atoms and (h) Co and Ru atoms in 1T-Ti$_{0.25}$V$_{0.25}$Co$_{0.25}$Ru$_{0.25}$S$_{2}$, with the total DOS shown by the black curves. Panels (i–l) show the corresponding DOS plots for 1H-Ti$_{0.25}$Co$_{0.25}$Mo$_{0.25}$Ru$_{0.25}$Se$_{2}$ and 1T-Ti$_{0.25}$V$_{0.25}$Co$_{0.25}$Ru$_{0.25}$Se$_{2}$, respectively.
}
\label{fig:fig3}
\end{figure}

\begin{figure}[tbp]
\includegraphics[width=1.00\columnwidth,clip=true,angle=0]{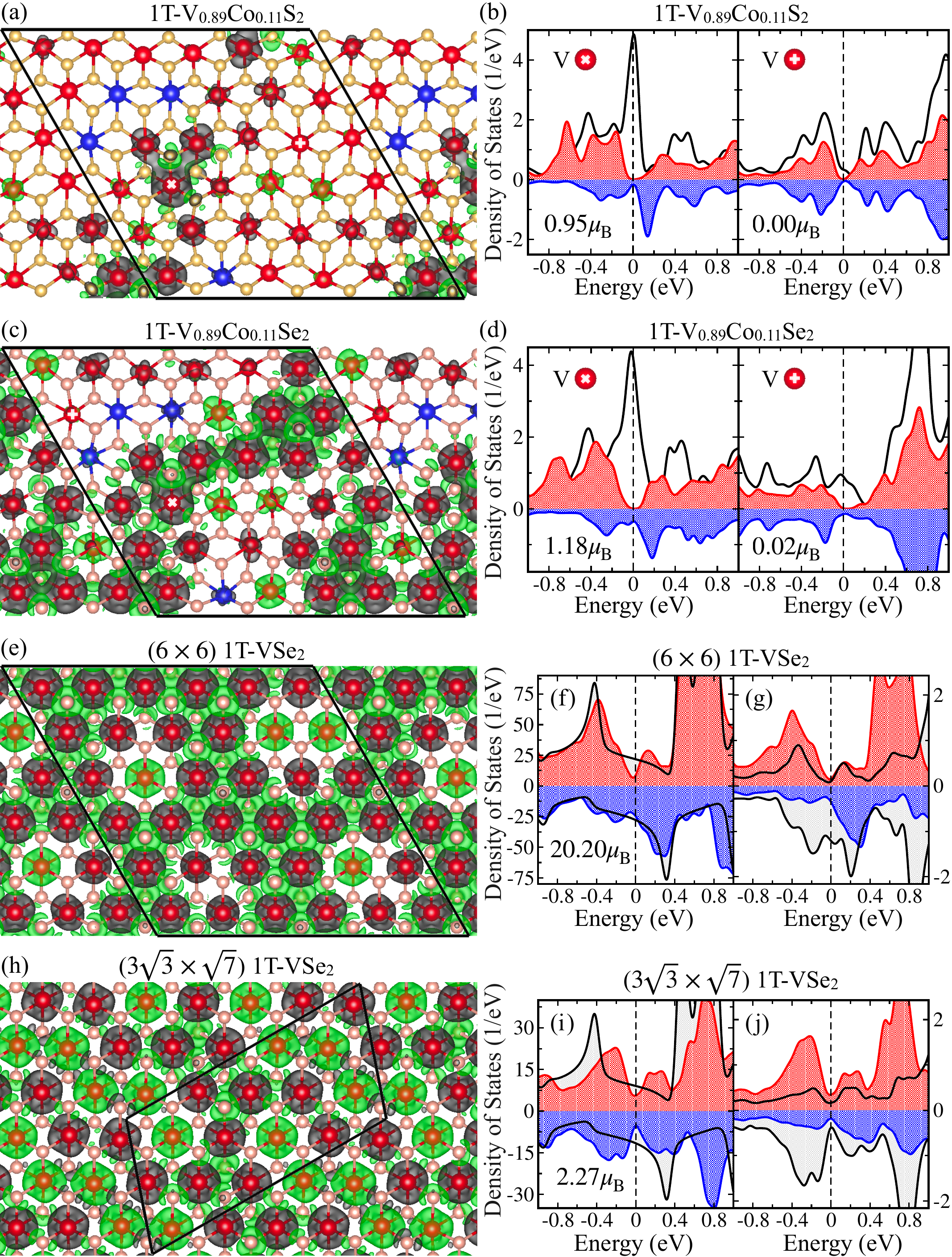}
\caption{(a) Spin density, and (b) spin-up (red) and spin-down (blue) DOS for selected atoms with their spin moments in 1T-V$_{0.89}$Co$_{0.11}$S$_2$ modeled with a $6\times6$ supercell. (c) Spin density and (d) DOS for 1T-V$_{0.89}$Co$_{0.11}$Se$_2$. Spin density for (e) ($6\times6$) 1T-VSe$_2$ and (h) ($3\sqrt{3}\times\sqrt{7}$) 1T-VSe$_2$, both containing V atoms with reversed spin moments, with the total DOS shown in (f) and (i). The nonmagnetic DOS (black curves), calculated using the same structures, is shown in (b), (d), (f), and (i) for comparison. (g) DOS averaged over four V atoms with the most positive spin moments (red and blue) and the most negative ones (black curves). (j) Averaged DOS from V atoms with positive spin moments (red and blue) and negative ones (black curves).
Positive and negative spin densities are shown in black and green, respectively, with the isovalue of the isosurfaces set to 0.002 electron/Bohr$^3$.
}
\label{fig:fig4}
\end{figure}

The sensitivity of the DOS at the Fermi level to chalcogen height suggests that mixing nonmagnetic X(S,Se)$_2$ compounds may induce magnetism around certain X sites if local structures yield a sufficiently high DOS at the Fermi level. Conversely, mixing magnetic V(S,Se)$_2$ with other compounds may suppress magnetism. Mixed X(S,Se)$_2$ compounds thus provide a rich platform for exploring diverse magnetic structures. Introducing disorder is generally expected to increase the total energy. However, our previous study found that mixing monolayer 1H-XS$_2$ compounds can lower the total energy and eliminate phonon instabilities\cite{Lee}. Because the unstable vibrational modes vary across the unmixed compounds, altering the elemental composition can stabilize otherwise unstable modes. This stabilization is associated with charge transfer and the development of CDWs. Thus, even when each unmixed compound is free of phonon instabilities in a sufficiently large supercell, the mixed phase may still have a lower total energy. We expect similar behavior in mixed 1H-XSe$_2$ and 1T-X(S,Se)$_2$.
Figs.~\ref{fig:fig3} (a)-(d) show the spin densities of four dynamically stable mixed phases. 1H-Ti$_{0.25}$Co$_{0.25}$Mo$_{0.25}$Ru$_{0.25}$(S,Se)$_2$ exhibits magnetism, whereas its unmixed counterparts are nonmagnetic. In contrast, 1T-Ti$_{0.25}$V$_{0.25}$Co$_{0.25}$Ru$_{0.25}$(S,Se)$_2$ has zero spin moments, even though 1T-V(S,Se)$_2$ is ferromagnetic. This can again be attributed to Stoner instability, as shown by comparing spin-unpolarized and spin-polarized DOS in Figs.~\ref{fig:fig3} (e), (f), (i), and (j), and there are no peaks at the Fermi level in Figs.~\ref{fig:fig3} (g), (h), (k), and (l). As reported in Table~\ref{table:totalenergy}, all four mixed compounds have negative mixing energies, indicating greater energetic stability than their unmixed counterparts in the same 1H or 1T structure. Furthermore, the negative relative energies of 1T-Ti$_{0.25}$V$_{0.25}$Co$_{0.25}$Ru$_{0.25}$(S,Se)$_2$ indicate that these mixed phases are energetically favored over the corresponding unmixed 1H and 1T structures, suggesting that 1T-V(S,Se)$_2$ may experience greater energy lowering with impurities than 1H-V(S,Se)$_2$. We speculate that this may explain why 1T-V(S,Se)$_2$ is more readily observed experimentally, even though its 1H form generally has a lower energy in the absence of impurities\cite{Zhang,Li,Esters,Luo}.

We now examine how introducing Co impurities alone can suppress ferromagnetism in 1T-VS$_2$. As shown in Fig.~\ref{fig:fig4} (a), the spin density vanishes near the Co atoms. This influence extends to more distant V atoms, such as the atom marked with a plus sign, whose spin moment is reduced to nearly zero. Fig.~\ref{fig:fig4} (b) shows that the spin-unpolarized DOS at the Fermi level varies between V atoms, with higher DOS associated with greater spin splitting in spin-polarized calculations. These results suggest that Co impurities can disrupt ferromagnetism. For the same structural arrangement, 1T-V$_{0.89}$Co$_{0.11}$Se$_2$ exhibits weaker magnetic suppression, as shown in Fig.~\ref{fig:fig4} (c). Nevertheless, the same relationship between a higher spin-unpolarized DOS at the Fermi level and greater spin splitting persists, as illustrated in Fig.~\ref{fig:fig4} (d). Despite increased structural and magnetic disorder, the 1T-V$_{0.89}$Co$_{0.11}$(S,Se)$_2$ remain dynamically stable and have negative mixing energies, as reported in Table~\ref{table:totalenergy}.

Magnetic disorder can also be introduced by reversing the local spin moments on a subset of V atoms. Figs.~\ref{fig:fig4} (e) and (h) illustrate two ferrimagnetic phases of 1T-VSe$_2$ in (6$\times$6) and ($3\sqrt{3}\times\sqrt{7}$) supercells, respectively. Previous studies have explored antiferromagnetic and ferrimagnetic phases of 1T-VSe$_2$\cite{Fuh,Esters,Yin}; however, the two dynamically stable phases identified here have mixing energies of -0.2 meV/atom (see Table~\ref{table:totalenergy}), making them nearly degenerate with ferromagnetic 1T-VSe$_2$. More energetically degenerate ferromagnetic, ferrimagnetic, and antiferromagnetic states are expected across a variety of supercell sizes and shapes. The near degeneracy of these dynamically stable magnetic configurations indicates strong competition between magnetic orders. As shown in Figs.~\ref{fig:fig4} (f) and (i), the DOS splitting and suppression at the Fermi level, relative to the ferromagnetic case, reduce the energy penalty for magnetic and structural disorder. Because ferromagnetic 1T-VSe$_2$ exhibits no phonon instability, this DOS arrangement is of electronic origin. Further analysis of the DOS contributions from the ferromagnetically aligned V atoms and those with reversed local spin moments reveals that the reduction in the DOS at the Fermi level cannot be attributed solely to the latter, as shown in Figs.~\ref{fig:fig4} (g) and (j), suggesting a collective electronic response.

In conclusion, mixing nonmagnetic X(S,Se)$_2$ compounds may induce magnetism, whereas mixing magnetic V(S,Se)$_2$ with other compounds may suppress it entirely. Such disorder-mediated magnetism may further lower the total energy, making mixed X(S,Se)$_2$ compounds a rich platform for exploring diverse magnetic structures. In 1T-VSe$_2$, ferromagnetism can be disrupted by impurities or by the formation of ferrimagnetic patterns at no energy cost. Therefore, the ground-state 1T-VSe$_2$ is not simply ferromagnetic. The energetic degeneracy between the ferromagnetic and ferrimagnetic states and their competition offer a mechanism for suppressing long-range ferromagnetic order. The reduced DOS at the Fermi level in ferrimagnetic 1T-VSe$_2$ is consistent with behavior associated with Stoner and Peierls instabilities, suggesting that energy lowering driven by electronic states near the Fermi level may be widespread among TMD monolayers.

\begin{acknowledgments}
The calculations were carried out using the computational facilities at JAIST and Tamkang University. 
C.-C. Lee acknowledges financial support from the National Science and Technology Council of Taiwan under Grant No. NSTC 114-2112-M-032-014, 
as well as support from the National Center for Theoretical Sciences of Taiwan.
\end{acknowledgments}

\bibliography{refs}

\end{document}